# An Improved Approach for Output Feedback Model Predictive Control of Hybrid Systems

**Soroush Sadeghnejad**[*,1], **Farshad Khadivar**[2], **Mojtaba Esfandiari**[2], **Golchehr Amirkhani**[2,]

**Hamed Moradi**[2],

**Farzam Farahmand**[2], **Gholamreza Vossoughi**[2]

[1]Biomedical Engineering Department, Amirkabir University of Technology (Tehran Polytechnic). No. 350, Hafez Ave, Valiasr Square, Tehran 1591634311, I.R. IRAN.

[2]Mechanical Engineering Department, Sharif University of Technology. Azadi Ave., Tehran 1458889694, I.R. IRAN.

*Corresponding Author
Assistant Professor, Biomedical Engineering Department, Amirkabir University of Technology (Tehran Polytechnic)
Email: s.sadeghnejad@aut.ac.ir
Tel: +98 (21) 64545952; Fax: +98 (21) 66468186

**Abstract:** In this paper, a modified robust model predictive control scheme is proposed for linear parametric variable (LPV) and hybrid systems based on a quasi-min-max algorithm. Using a new cost function resulted in reduced unwanted disturbances during switching. In addition, the effects of uncertainties are reduced in the prediction dynamics, and robust stability of piecewise switching LPV systems subjected to linear matrix inequality (LMI) input constraints are guaranteed. Simulation results demonstrate the performance of the proposed controller compared to prior research.

## 1. INTRODUCTION

It is well known that linear dynamic models cannot be used generally for proper modeling of nonlinear characteristics existing in most physical systems [1-3]. Although there are many nonlinear control methods for nonlinear dynamic systems, they are computationally difficult to be implemented [4-6]. Using piecewise linear approaches is a common method to implement standard procedures and reduce the computational costs for the control of nonlinear systems [7-9]. But, due to the switching phenomenon in the dynamics of piecewise linear systems, the control signals may not be continuous and thus abruptly change the behavior of the system. Additionally, controlling a system with constraints on the control input, control output, and state variables is one of the challenging problems in the control theory, especially when robustness against uncertainties is required.

Model Predictive Control (MPC) is one of the advanced and novel types of control approaches; once known as Dynamic Matrix Control (DMC), developed in late 1976 to solve multivariable constrained problems [10-12]. It has been widely used in the control of several industrial processes, such as refining, petrochemicals, air and gas industries, automotive, mining/metallurgy, spacecraft, as well as robotics [13-15]. The use of MPC has been increased because it can handle the nonlinear multiple-inputs multiple-outputs (MIMO) systems with switching phenomena and constraints on inputs and outputs [16-18]. However, as with any other model-based control technique, MPC requires a precise dynamic model and full-state estimation.

The study of MPC approaches, that guarantee the stability of the system as well as the satisfaction of all constraints, is an attractive research area [19-21]. In an early study, minimax formulations for robust MPC were investigated by Campo and Morari [22] and Raimondo et al. [23]. In [24], Kothare et al. used semi-definite programming and linear matrix inequalities (LMIs) for robust constrained MPC. Similarly, a state feedback MPC scheme is proposed by Lu and Arkun [25]. Using a quasi-min-max algorithm, they modeled the uncertainties in both linear time-varying (LTV) and linear time-invariant (LTI) systems. In recent studies, one can find different approaches based on state feedback methods, where the system states are assumed to be measurable [26-28].

Having difficulties in guaranteeing the stability of the closed-loop control system, some researchers considered the output feedback MPC approaches to stabilize the system [29]. There are fewer output feedback MPC algorithms in comparison with state feedback MPC algorithms in the literature. In [30], Wan and Kothare expanded their previously proposed method [24] into an off-line robust output feedback MPC for a wide range of constrained LTV systems. In a recent study, Park et al. [31] improved the proposed method [30] and designed an online robust output feedback MPC algorithm for linear parameter varying (LPV) systems based on a quasi-min-max algorithm. The proposed method guaranteed the robust stability of the output feedback systems with constraints. However, the lack of smooth behaviour during any transition between piecewise linear models of the system is the main disadvantage of these methods. Therefore, while it is necessary to have continuous controls in many cases, the available control approaches are unable to satisfy this condition. Moreover, the designed control system in the mentioned work has a non-zero initial value which causes implementation problems in practice.

In this paper, an improved approach for quasi-min-max output feedback model predictive control for implementation on linear parameter variable (LPV), as well as hybrid systems, is proposed. A modified robust output feedback MPC scheme for LPV systems based on a quasi-min-max algorithm is developed to reduce the unwanted disturbances caused by switching in the dynamics of the control effort signals. Besides, the effect of available uncertainties in the prediction dynamics is reduced, resulting in an improvement in the system's robustness and the convergence of the output signals to the desired values. For developing this new approach, some assumptions are made similar to the work by Park et al. [31]. It is assumed that the current system parameters are measurable. Our proposed framework makes the following novel contributions:

- ✓ In contrast to the previous approaches [31] where the next state $x_{k+1}$ is considered as an unknown system's future state, our approach utilizes the predictive dynamics. So, it reduces the entire system's uncertainties and improves output convergence.
- ✓ The definition of the optimization cost function is changed and $\Delta u$ is utilized instead of $u$. So, it is possible to not only eliminate the unwanted disturbances caused by switching and discontinuities in the control law but also to guarantee that the initial value of the control signal remains zero at $t = 0$.

The aforementioned contributions and characteristics of the proposed approach are novel in the area of quasi-min-max output feedback MPC methods. Through an empirical evaluation, it is shown that the proposed modified MPC scheme for linear parametric variable or hybrid systems can guarantee the robust stability of the output feedback systems subject to the input constraints in a more practical aspect.

Based on the research [31], we first proposed an off-line robust state observer for LPV systems, using linear matrix inequality (LMI) [32]. Then, it is shown how to modify and improve the online optimization problem for the robust output feedback MPC scheme with LMI constraints. The proposed novelties will be compared with the system suggested by [31] and it is observed that the output performance is preserved while the control process has been improved to be more practical. In general, the proposed modified MPC scheme for LPV systems, subject to input constraints, can guarantee the robust stability of the output feedback systems. The new contributing features of the proposed modified MPC scheme are investigated through a numerical example. In the mentioned example, the proposed algorithm is implemented on a hybrid system for two different scenarios, one with defined and the other with random switching signals.

## 2. PROBLEM STATEMENT

### 2.1. Description of the Dynamic System

Consider a MIMO linear parameter variable (LPV) discrete-time system, described by the following dynamic model:

$$x(k+1) = A(s(k))x(k) + B(s(k))u(k) \tag{1}$$

$$y(k) = C(s(k))x(k)$$

in which, $x(k) \in R^{n_x}$ is the unmeasurable state of the system, $y(k) \in R^{n_y}$ is the measurable output, $u(k) \in R^{n_u}$ is the control input, and $s(k)$ is the time-varying switching signal which is measurable at each step $k$, and can be defined either dependent on or independent of the dynamic system. It is assumed that $[A(s(k)), B(s(k)), C(s(k))] \in \Omega \equiv C_0\{[A_i, B_i, C_i]\}$ varies inside a corresponding polytope $\Omega$ whose vertices consist of $L_g$ local system matrices (number of piece-wise linear regions) and are known at each step $k$. Besides, $C_0$ denotes the convex hull of the aforementioned polytope.

## 2.2 Design of Off-Line Robust Observer

In the previous section, the dynamics of an LPV discrete-time system was introduced. Given that the states of the system are unmeasurable, an off-line robust observer is used based on [31]. The following state observer is proposed to estimate the uncertainties in the state of the system in Eq. (1), as:

$$\hat{x}(k+1) = A(s(k))\hat{x}(k) + B(s(k))u(k) + L_o\left(y(k) - C(s(k))\hat{x}(k)\right) \tag{2}$$

where, $\hat{x}(k)$ is the estimated state of the system and $L_o$ is the observer gain which is necessary to be determined in terms of LMIs.

Park et al. [31] designed an off-line robust observer in order to estimate the states of the dynamic system from its output feedback. In this regard, the estimation error of states is defined as $e_s(k) = x(k) - \hat{x}(k)$. Considering Eqs. (1) and (2), the error dynamics can be defined as:

$$e(k+1) = \left(A(s(k)) - L_0 C(s(k))\right)e(k) \tag{3}$$

Consider a quadratic function for the estimation error $e(k)$, as $E(e(k))$,

$$E(e(k)) = e(k)^T P_0 e(k), \quad P_0 > 0 \tag{4}$$

in which, $P_0$ is a positive definite matrix. If $E(e(k))$ satisfies the following condition:

$$E(e(k+1+i)) - \rho^2 E(e(k+i)) < -e(k+i)^T L_0 e(k+i) \tag{5}$$

then it is possible to guarantee that $e(k) \to 0$ as $k \to \infty$. In Eq. (5), for $i \geq 0$, $\rho$ and $L_0$ are the decay rate and the weighting matrix respectively and are determined by the designer. Substituting $\Lambda(k+i) = A(s(k+i)) - L_0 C(s(k+i))$ in Eq. (5), the equation can be rewritten as follow:

$$\rho^2 P_0 - L_0 - \Lambda(k+i)^T P_0 \Lambda(k+i) > 0 \tag{6}$$

Using the Schur complement [33], Eq. (6) is explained as:

$$\rho^2 P_0 - L_0 - (P_0 \Lambda(k+i))^T P_0^{-1} (P_0 \Lambda(k+i)) > 0 \tag{7}$$

and by replacing $Y_0 = P_0 L_0$, the following inequality is obtained:

$$\begin{bmatrix} \rho^2 P_0 - L_0 & * \\ P_0 A_j - Y_0 C & P_0 \end{bmatrix} > 0 \tag{8}$$

where * denotes the transpose of $P_0 A_j - Y_0 C$. Given the inequality defined by Eq. (8) is affine in $\left[\left(A(s(k+i))\middle| B(s(k+i))\right)\right]$, and satisfying the input constraint; then $\left[\left(A(s(k+i))\middle| B(s(k+i))\right)\right] \in \Omega$, if and only if there exist appropriate $Y_0$ and $P_0 > 0$. Thus the observer gain $L_0$ is obtained as $L_0 = P_0^{-1} Y_0$ and hence the estimated state vector $\hat{x}(k)$ in Eq. (2) converges to the real states of the system $x(k)$.

**In summary**, if there exists a $P_0 > 0$ and $Y_0 = P_0 L_0$ which satisfies the constraint of Eq. (8), then the obtained observer gain $L_0$ and the stability of the dynamic error of Eq. (4) can be guaranteed for any $\left[\left(A(s(k))\middle| B(s(k))\right)\right] \in \Omega$ and finally, as $k \to \infty$, one can conclude that $\hat{x}(k) \to x(k)$.

### 2.3. MPC-Based Control with On-Line Output Feedback

To achieve the desired control function for a dynamic system (e.g., regulation of the output signal), it is required first to define the cost function and the constraints of the system. Afterward, based on our assumptions and concepts given in section 2.2, the objective is to design a control scheme that can stabilize the system through optimization of the cost function while satisfying the associated constraints. Given the dynamics of the system in Eq. (1), it is assumed that the current system matrices $\left[\left(A(s(k))\middle| B(s(k))\right)\right]$ are known at step $k$. At the same time, the future ones $\left[\left(A(s(k+i))\middle| B(s(k+i))\right)\right]$ for $i \geq 1$ is unknown as to the defined polytope $(\Omega)$. It is assumed that the system input is constrained as:

$$|u_j(k)| < u_{j,max} \quad j = 1,2,\ldots,n_u \quad k \geq 0 \tag{9}$$

where, $|.|$ is the absolute value, and $u_{j,max}$ denotes the given upper bound on the $j^{th}$ input $u_j(k)$. Since the states of Eq. (1) are immeasurable, the observer proposed in section 2.2 is considered.

Park et al. [31] and Lu and Arkun [25], have proposed different online/offline state feedback model predictive control strategies for LPV systems in conjunction with state observers. In research [25], at each step, a minimization occurs on the "quasi-worst-case" value of a quadratic objective function in an infinite horizon.

Since in the theories introduced by Lu and Arkun [25] and Park et al. [31], no future step is calculated; the predictive dynamics is not used to its full capability. However, as the current state of the dynamic system is available either measured or estimated, the predictive dynamic model is able to predict the dynamics of the system one or more steps into the future. Therefore, the uncertainties associated with the prediction and optimizations are reduced. Hence, one of the main contributions of this research is to reduce the uncertainties in the predicted dynamics at the next iteration. Moreover, another contribution of this research is a new cost function by utilizing $\Delta u$ rather than $u$. This contribution leads to the elimination of unwanted disturbances caused by switches and discontinuities in the control efforts. In addition, it guarantees the initial value of the control signal to be zero at $t = 0$, which is important in practice. Given this control scheme, the prediction conservation is reduced while satisfying the constraints.

In our model predictive control, in order to predict the future states of the system presented in Eq. (1), the following predictive model is introduced:

$$\hat{x}_p(k + 1 + i|k) = A(s(k + i))\hat{x}_p(k + i|k) + B(s(k + i))u(k + i|k)$$
$$\hat{x}_p(k|k) = \hat{x}(k), \quad i \geq 1 \tag{10}$$

where, $\hat{x}_p(k + i|k)$ denotes the predicted state of $x(k + i)$ of the system in Eq. (1) and $u(k)$ is the future control input for step $k + i$ considered at step $k$. For this purpose, assume that $\hat{x}_p(k|k) := \hat{x}(k)$ and the first element of the computed control sequence is $u(k|k)$ and in the future steps, $u(k + i|k)$ is estimated as follows:

$$u(k + i|k) = \Psi(k)\hat{x}_p(k + i|k), \quad i \geq 1 \tag{11}$$

where, $\Psi(k)$ is the future feedback gain introduced at step $k$. Using the optimization strategy introduced in [31], this method proposes to minimize the infinite horizon objective function $J_0^\infty(k)$ at each step $k$ as:

$$\min_{u(k + i|k)} \max_{[(A(s(k + i))|B(s(k + i)))] \in \Omega, i \geq 0} J_0^\infty(k)$$
$$J_0^\infty(k) = J_0^1(k) + J_1^\infty(k) \tag{12}$$

Equation (12) should be supplemented with the input constraints in (9). In (12), the terms are represented as:

$$\begin{cases} \Delta u(k) = u(k) & k = 1 \\ \Delta u(k) = u(k) - u(k - 1) & k > 1 \end{cases}$$

$$J_0^\infty(k) = \sum_{i=0}^{\infty}\{\hat{x}_p(k + i|k)^T Q\hat{x}_p(k + i|k) + \Delta u(k + i|k)^T R\Delta u(k + i|k)\}$$

$$J_0^1(k) := \hat{x}_p(k|k)^T Q\hat{x}_p(k|k) + \Delta u(k|k)^T R\Delta u(k|k) \tag{13}$$

$$J_1^\infty(k) = \sum_{i=1}^{\infty}\{\hat{x}_p(k + i|k)^T Q\hat{x}_p(k + i|k) + \Delta u(k + i|k)^T R\Delta u(k + i|k)\}$$

$$J_1^\infty(k) := \sum_{i=1}^{\infty} \{\hat{x}_p(k+i|k)^T Q \hat{x}_p(k+i|k) + \Delta u(k+i|k)^T R \Delta u(k+i|k)\}$$

$$= \hat{x}_p(k+1|k)^T Q \hat{x}_p(k+1|k) + J_2^\infty$$

$$J_2^\infty = \sum_{i=2}^{\infty} \{\hat{x}_p(k+i|k)^T Q \hat{x}_p(k+i|k)\} + \sum_{i=1}^{\infty} \{\Delta u(k+i|k)^T R \Delta u(k+i|k)\}$$

where $R > 0$ and $Q > 0$ are the weight matrices. By introducing an upper bound on $J_2^\infty$, which is different from what is proposed in [25, 31], the quasi-min–max optimization problem for MPC can be derived. By defining the following Switched Quadratic Lyapunov function:

$$V\left(\hat{x}_p(k+i|k)\right) = \hat{x}_p(k+i|k)^T \Gamma(k) \hat{x}_p(k+i|k), \quad i > 0, \Gamma(k) > 0 \tag{14}$$

it is assumed that $V\left(\hat{x}_p(k+i|k)\right)$ satisfies the following stability condition at each step k for $\left[\left(A(s(k+i))\big|B(s(k+i))\right)\right] \in \Omega, i \geq 1$. Then, (14) can be rewritten as:

$$V_{k+i} = Xp_{k+i} \Gamma_k Xp_{k+i} \tag{15}$$

Equation (15) holds for any future step that results in the folowing relation,

$$V_{k+i+1} - V_{k+i} < -J_{k+i}^{k+i+1}$$
$$J_{k+i}^{k+i+1} = Xp_{k+i} Q Xp_{k+i} + \Delta u_{k+i}^T R \Delta u_{k+i} \tag{16}$$

To guarantee the stability of the system, it is necessary that $V\left(\hat{x}_p(\infty|k)\right) = 0$. In addition, it is necessary to optimize all cost functions even in all future steps. Therefore, the stability criterion in Eq. (16) for each future step is added to the criterion of the other steps in order to achieve an upper bound for the summation of all cost functions based on $V_k$ and $V_{k+1}$. Then, the resulting control law will be stable at the current and any future time step. Using $V(k)$, it can be defined:

$$J_0^\infty(k) < V_k \tag{17}$$

where

$$J_2^\infty + Xp_{k+1}^T Q Xp_{k+1} + J_0^1 < V_k \tag{18}$$

or

$$J_2^\infty < V_k - Xp_{k+1}^T Q Xp_{k+1} - J_0^1 \tag{19}$$

In (19), the left side of the inequality is indefinite and the right side shows the maximum of $J_2^\infty$. Optimizing the cost function while satisfying the input constraints, the control input is defined as:

$$\max J_2^\infty < \delta(k) \tag{20}$$

where, $\delta(k)$ is a non-negative variable to be minimized. Considering (20) for each time step, the following relation can be derived:

$$V_k - Xp_{k+1}^T Q Xp_{k+1} - J_0^1 < \delta_k \tag{21}$$

If $\Gamma(k) > \varepsilon I$ for $\varepsilon > 0$ is considered, then the additional constraint will guarantee the stability of the system in Eq. (18) [31]. In summary, to guarantee the stability of an LPV system with input constraints using MPC, the quasi-min–max optimization problem of Eq. (12) must be solved in order to find the $u(k|k)$ as the control input and $\Psi(k)$ as the future feedback gain. Henceforth, we focus on the design of the MPC scheme for an LPV system with LMI constrints and solve the quasi-min-max optimization problem of the infinite horizon MPC scheme. This is synonymous to find the solutions for $u(k) \coloneqq u(k|k)$ and $\Psi(k)$ by solving the optimization problem in Eq. (12) while guaranteeing the stability of the system and including the effects of constraints on the performance and input variables. Hence, the following functions can be defined:

$$V_k = Xp_k^T \Gamma_k Xp_k$$
$$Xp_{k+1} = A_k Xp_k + B_k \Delta u(k) + L_o(y(k) - C_k \widehat{X}_k) + u(k-1) \tag{22}$$
$$J_0^1 = Xp_k^T Q Xp_k + \Delta u_k^T R \Delta u_k$$

Cnsidering Eqs. (21) and (22), Eq. (21) can be rewritten as:

$$Xp_k^T \Gamma_k Xp_k - (Xp_k^T + Xp_k^T A_k^T) Q (Xp_k + A_k Xp_k) - \Delta u^T (B^T Q B) \Delta u - \Delta u^T R \Delta u$$
$$- [L_o(y - C_k \widehat{X}_k) + u(k-1)]^T Q [L_o(y - C_k \widehat{X}_k) + u(k-1)] < \delta_k \tag{23}$$

in which it is supposed that:

$$\delta \tau_k = (Xp_k^T + Xp_k^T A_k^T) Q (Xp_k + A_k Xp_k) + \Delta u^T (B^T Q B + R) \Delta u \tag{24}$$
$$+ [L_o(y - C_k \widehat{X} p_k) + u(k-1)]^T Q [L_o(y - C_k \widehat{X}_k) + u(k-1)]$$

Then, the following inequality can be derived:

$$1 + \tau_k - Xp_k^T \Phi^{-1} Xp_k > 0 \tag{25}$$

Using the Schur complement, Eq. (25) results in:

$$\begin{bmatrix} 1 + \tau_k & * \\ Xp_k & \Phi(k) \end{bmatrix} > 0 \tag{26}$$

Defining the complementary equation and some auxiliary matrices as:

$$\begin{bmatrix} \eta & * \\ Xp_k & \Phi(k) \end{bmatrix} + \Lambda^T \Phi^{-1} \Lambda > 0$$
$$\Lambda = \begin{bmatrix} \theta^{\frac{1}{2}} Xp_k & 0 \\ \beta^{\frac{1}{2}} \Delta u_k & 0 \end{bmatrix}, \quad \begin{cases} \theta = (I + A^T) Q (I + A) \\ \beta = B^T Q B + R \end{cases}, \tag{27}$$
$$\eta = 1 + [L_o(y - C_k \widehat{X}_k) + u(k-1)]^T Q [L_o(y - C_k \widehat{X}_k) + u(k-1)]$$

Then, again using the Schur complement, the aforementioned inequality will result in:

$$\begin{bmatrix} M & * \\ T & \Pi \end{bmatrix} > 0 \tag{28}$$

$$M = \begin{bmatrix} \eta & * \\ Xp_k & \Phi \end{bmatrix}, \quad T = \begin{bmatrix} \theta^{\frac{1}{2}} Xp_k & 0 \\ \beta^{\frac{1}{2}} u_k & 0 \end{bmatrix}, \quad \Pi = \begin{bmatrix} \delta I & 0 \\ 0 & \delta I \end{bmatrix}$$

Hence, for the next step, it is necessary to define a constraint which can guarantee the stability of the system with regard to the future inputs. Substituting $u(k + i|k) = \Psi(k)\hat{x}_p(k + i|k)$, $i \geq 1$, and also considering Eq. (15) and the future states of the prediction model in Eq. (10), yields:

$$\hat{x}_p(k + i|k)^T \mu(k)\hat{x}_p(k + i|k) - \hat{x}_p(k + i|k)^T \omega(k + i)^T \Gamma(k)\omega(k + i)\hat{x}_p(k + i|k) > 0 \quad (29)$$

where,

$$\mu(k) = \Gamma(k) - Q - \Psi(k)^T R \Psi(k)$$
$$\omega(k + i) = A(s(k + i)) + B(s(k + i))\Psi(k) \quad (30)$$

Therefore, the following relation can be derived such that:

$$\mu(k) - \omega(k + i)^T \Gamma(k)\omega(k + i) > 0 \quad (31)$$

Multiplying by $\delta(k)$ and pre- and post-multiplying by $\Gamma(k)^{-1}$, this inequality becomes:

$$\delta(k)\left(\Gamma(k)^{-1} - (\Gamma(k)^{-1})^T Q(\Gamma(k)^{-1}) - (\Psi(k)\Gamma(k)^{-1})^T \times R(\Psi(k)\Gamma(k)^{-1})\right)$$
$$- (\delta(k)\Gamma(k)^{-1})^T \omega(k + i)^T \delta(k)^{-1} \times \Gamma(k)\omega(k + i)(\delta(k)\Gamma(k)^{-1}) \quad (32)$$

Considering $\Gamma(k) = \delta(k)\Phi(k)^{-1}$ and again using the Schur complement, Eq. (32) can be shown to be equivalent to:

$$\begin{bmatrix} N(k) & * \\ \omega(k + i)\Phi(k) & \Phi(k) \end{bmatrix} > 0 \quad (33)$$

where,

$$N(k) = \delta(k)(\Gamma(k)^{-1} - (\Gamma(k)^{-1})^T Q(\Gamma(k)^{-1}) - (\Psi(k)\Gamma(k)^{-1})^T R(\Psi(k)\Gamma(k)^{-1}) \quad (34)$$

By assuming $Y(k) = \delta(k)\Phi(k)$, Eq. (33) will result in:

$$\begin{bmatrix} \Phi(k) & * \\ S(k) & \Phi(k) \end{bmatrix} - \Theta(k)^T \Pi(k)\Theta(k) > 0 \quad (35)$$

where,

$$S(k) = A(s(k + i))\Phi(k) + B(s(k + i))Y(k)$$
$$\Theta(k) = \begin{bmatrix} Q^{\frac{1}{2}}\Phi(k) & 0 \\ R^{\frac{1}{2}}Y(K) & 0 \end{bmatrix} \quad (36)$$

In addition, using the Schur complement, the following matrix inequality can be obtained:

$$\begin{bmatrix} \Phi(k) & * & * & * \\ S(k) & \Phi(k) & * & * \\ Q^{1/2}\Phi(k) & 0 & \delta(k)I & * \\ R^{1/2}Y(k) & 0 & 0 & \delta(k)I \end{bmatrix} > 0 \quad (37)$$

where * denotes that the matrix is axisymmetric. The aforementioned inequality by Eq. (37) is affine in $\left[\left(A(s(k + i))\big|B(s(k + i))\right)\right]$ if and only if $Y(k)$ and $\Phi(k)$ exist such that:

$$\begin{bmatrix} \Phi(k) & * & * & * \\ S_j(k) & \Phi(k) & * & * \\ Q^{1/2}\Phi(k) & 0 & \delta(k)I & * \\ R^{1/2}Y(k) & 0 & 0 & \delta(k)I \end{bmatrix} > 0, \quad j = 1, 2, \ldots, L_g \tag{38}$$

where, $S_j(k) = A_j\Phi(k) + B_jY(k)$ and the general format of the matrix is axisymmetric. Then, it can be denoted that Eq. (21) is satisfied for $\left[\left(A(s(k+i))\middle|B(s(k+i))\right)\right] \in \Omega$. Considering $\Gamma(k) = \delta(k)\Phi(k)^{-1}$ for some $\epsilon > 0$, it results in:

$$\delta(k)I - \epsilon\Phi(k) > 0 \tag{39}$$

In addition, it is necessary to consider the constraints on the control input and also for the future control sequences $u(k + i|k)$. For those constraints, it is required to define an upper bound for the control input as follows:

$$|u(k + i|k)| < u_{j,max}, \quad j = 1, 2, \ldots, n_u, \quad i \geq 1 \tag{40}$$

Hence, it can be concluded that:

$$\max_{i \geq 1}|u(k+i|k)|^2 = \max_{i \geq 1}\left|\left(Y\Phi^{-1}\hat{z}(k+i|k)\right)_j\right|^2 \leq \max_{i \geq 1}\left|(Y\Phi^{-1}\hat{z})_j\right|^2 \tag{41}$$

$$\leq \left\|\left(Y\Phi^{-1/2}\right)_j\right\|_2^2 = (Y\Phi^{-1}Y^T)_{jj}$$

Therefore, if there exists a symmetric matrix, $Su$, such that:

$$\begin{bmatrix} Su(k) & * \\ Y(k)^T & \Phi(k) \end{bmatrix} > 0 \wedge Su_{jj} < u_{j,max}^2 \quad j = 1, 2, \ldots, n_u \tag{42}$$

Then, it guarantees that:

$$|u(k + i|k)| < u_{j,max}, \quad j = 1, 2, \ldots, n_u, \quad i \geq 1 \tag{43}$$

Therefore, the input constraint by Eq. (43) is guaranteed for, $j = 1, 2, \ldots, n_u, \quad i \geq 1$.

Theorem: For a linear parameter variable system in Eq. (1), which is constrained by Eq. (9), the optimization problem in Eq. (12) will result in finding the $u(k)$ and $\Psi(k)$ for the future control law $u(k + i|k) = \Psi(k)\hat{x}_p(k + i|k)$, in which the upper bound of $\delta(k)$ for the infinite horizon objective function $J_0^\infty(k)$ minimizes the optimization problem,

$$\min_{u(k|k),Y(k),\Phi(k)} \delta(k) \tag{44}$$

Implementing the aforementioned optimization problem, it can be concluded that:

$$\begin{bmatrix} \eta & * & * & * \\ Xp_k & \Phi(k) & * & * \\ ((I+A)^TQ(I+A))^{1/2}\hat{x}(k) & 0 & \delta(k)I_{nx*nx} & * \\ (B^TQB+R)^{1/2}\Delta u(k) & 0 & 0 & \delta(k)I_{nu*nu} \end{bmatrix} > 0 \tag{45}$$

$$\begin{bmatrix} \Phi(k) & * & * & * \\ S_j(k) & \Phi(k) & * & * \\ Q^{1/2}\Phi(k) & 0 & \delta(k)I_{nx*nx} & * \\ R^{1/2}Y(k) & 0 & 0 & \delta(k)I_{nu*nu} \end{bmatrix} > 0, \quad j = 1,2,\ldots,L_g \tag{46}$$

$$\delta(k)I - \varepsilon\Phi(k) > 0 \tag{47}$$

$$\begin{bmatrix} u_{j,max} & u_j(k) \\ u_j(k) & u_{j,max} \end{bmatrix} > 0 \tag{48}$$

$$\begin{bmatrix} Su(k) & * \\ Y(k)^T & \Phi(k) \end{bmatrix} > 0 \wedge Su_{jj} < u_{j,max}^2 \quad j = 1,2,\ldots,n_u \tag{49}$$

in which, Eqs. (45) and (46) are axisymmetric matrixes and,

$$\begin{aligned} \eta &= 1 + \left[L_o(y - C_k\hat{X}_k)\right]^T Q\left[L_o(y - C_k\hat{X}_k)\right] \\ S_j(k) &:= A_j\Phi(k) + B_jY(k) \end{aligned} \tag{50}$$

If $u(k|k)$, $Y(k)$ and $\Phi(k)$ are available for $\hat{x}(k)$ and $y(k)$, the future feedback gain $\Psi(k)$ results from $\Psi(k) := Y(k)\Phi(k)^{-1}$. Eq. (45) to Eq. (50) are derived from the conditions and inequalities caused by Eq. (21), the stability criterion in Eq. (16), and $\Gamma(k) > \varepsilon I$. If there exist $Y(k)$, $\Phi(k)$ and $\delta(k)$ in somehow that Eq. (44) will hold and if

$$\hat{x}_p(k|k)^T Q\hat{x}_p(k|k) + \Delta u(k|k)^T R\Delta u(k|k) + \hat{x}_p(k+i|k)^T \Gamma(k)\hat{x}_p(k+i|k) < \delta(k) \tag{51}$$

Then

$$\hat{x}_p(k+i|k)^T \Gamma(k)\hat{x}_p(k+i|k) < \delta(k) \tag{52}$$

Assuming that $Q \& R > 0$, considering Eq. (16), then the result will be

$$\begin{aligned} &\hat{x}_p(k+i+1|k)^T \Gamma(k)\hat{x}_p(k+i+1|k) - \hat{x}_p(k+i|k)^T \Gamma(k)\hat{x}_p(k+i|k) \\ &< -\left[\hat{x}_p(k|k)^T Q\hat{x}_p(k|k) + \Delta u(k|k)^T R\Delta u(k|k)\right], \quad i \geq 1 \end{aligned} \tag{53}$$

that is

$$\hat{x}_p(k+i+1|k)^T \Gamma(k)\hat{x}_p(k+i+1|k) < \hat{x}_p(k+i|k)^T \Gamma(k)\hat{x}_p(k+i|k), \quad i \geq 1 \tag{54}$$

Therefore, $\varepsilon := \{\hat{x}_p \in R^{n_x} | \hat{x}_p^T \Gamma(k)\hat{x}_p < \delta\}$ is an invariant ellipsoid for the predicted future states $\hat{x}_p(k+i|k)$, $i \geq 1$, of the uncertain system. Then, based on Eqs. (51)-(54), the input constraint can be cast as LMI constraints as presented in Eqs. (40)-(42). For garunteeing the robust stability of the system, first it is needed to instate that the optimization problem, provided in Eq. (44), at step k is a feasible solution to the problem at step k+1. This is formulated in the folowing eqaution

$$\begin{aligned} \Delta u(k+1|k+1) &= \Delta u(k+1|k) = \Psi(k)\hat{x}_p(k+i|k) \\ \Gamma(k+1) &= \Gamma(k) \end{aligned} \tag{55}$$

Then, it is necessary to check Eqs. (45), (47), and (48), since they include $s(k)$, $\hat{x}(k)$, $y(k)$, and

$u(k|k)$, that change at each step k. First, by the satisfaction of (49) at step k, $\Delta u(k+1|k)$ is feasible. Then, considering $\Delta u(k+1|k+1) = \Delta u(k+1|k) = \Psi(k)\hat{x}_p(k+i|k)$, it is clearly stated that (48) is satisfied at step k + 1. In the next step, (45) should be checked. Based on (45) and Eqs. (51)-(54), the following terms are substituted as $\Delta u(k+1|k+1) = \Delta u(k+1|k) = \Psi(k)\hat{x}_p(k+i|k)$ and $\Gamma(k+1) = \Gamma(k)$. Then, (45) can be rewritten as

$$\hat{x}_p(k+1|k)^T \zeta(k) \hat{x}_p(k+1|k)$$
$$- \hat{x}_p(k+1|k)^T \hat{\xi}(k+1)^T \Gamma(k) \hat{\xi}(k+1) \hat{x}_p(k+1|k) < 0 \qquad (56)$$

in which $\zeta(k) := \Gamma(k) - Q - F(k)^T R F(k)$ and $\hat{\xi}(k+1) = A(s(k+1)) + B(s(k+1))F(k)$. Since at step k, Eq. (46) is satisfied, the presented inequality in Eq. (56) holds. Therefore, the feasibility of the optimization problem in Eq. (44) is guaranteed. Considering (46) and (47), it is guaranteed that there exists $\varepsilon > 0$ that satisfies $\delta(k)I - \varepsilon \Phi(k) > 0$ for any $t > k$. Now, the stability constraint (21) guarantees that $\hat{x}_p(k|k) := \hat{x}(k)$, $u(k|k)$, and $\hat{x}_p(k+1|k)$ converge to zero, since $Q \& R > 0$, and $\Gamma(k) > \varepsilon I$. In addition, it is concluded from section 2.2 that the estimated state $\hat{x}(k)$ will converge to the system state $x(k)$ as $k \to \infty$. Therefore, the feasibility of solving the optimization problem (44) guarantees the asymptotic robust stability, (i.e., $x(k) \to \infty$ as $k \to \infty$). Then, the proposed robust output feedback MPC scheme is described as $\min_{u(k|k),Y(k),\Phi(k),X(k)} \delta(k)$ subject to (10), (11), (45)–(49) and at each step k, for implementing the first control input $\Delta u(k|k)$ and $u(k|k)$, the optimal control sequence $u(k+i|k)$ will be determined by solving the aforementioned optimization, considering the measured output $y(k)$ and the time-varying parameter $s(k)$.

## 3. SIMULTION RESULTS

To demonstrate the effectiveness of the proposed quasi-min-max approach for output feedback model predictive control in comparison with the prior research [31], a numerical example is evaluated. The linear parametric variable system (LPV) switches between three different system dynamics which are given by:

$$x(k+1) = A_i(\alpha(k), \beta(k))x(k) + B_i(\gamma(k))u(k)$$
$$y(k) = C_i x(k) \qquad (57)$$

where,

$$A_i(\alpha(k), \beta(k)) = \begin{bmatrix} 0.8500 & -0.0743 \times \alpha_i(k) \\ 0.0811 \times \beta_i(k) & 0.9050 \end{bmatrix}$$
$$B_i(\gamma(k)) = \gamma_i(k) \begin{bmatrix} 0.1050 \\ 0.0092 \end{bmatrix}, \quad C_i = [0.65 \ -0.50] \quad i = 1,2,3 \qquad (58)$$

The unknown parameters are defined as $\alpha(k)\in\{1, 1.7, 2.4\}$, $\beta(k)\in\{1, 1.5, 4.3\}$ and

$\gamma(k)\epsilon\{0.33, 0.66, 1\}$, which are changed based on the switching signal either dependent on or independent of the operation point of the system. The initial states of the system are selected by Eq. (57) as $x(0) = [-1.5 \ -0.2]^T$ and the initial states of the observer are selected by Eq. (2) as $\hat{x}(0) = [0.5 \ 1]^T$. Taking into account the input constraint, introduced by Park et. al. [31], $|u(k)| < 1; k \geq 0$, and the decay rate; the weighting matrix and the state observer gain are obtained as given in Table 1.

Table 1 Observer parameters obtained from Theorem 1

| Parameters | Final value |
| --- | --- |
| Decay rate | $\rho = \sqrt{0.7}$ |
| Weighting matrix | $L_0 = \begin{bmatrix} 1 & 0 \\ 0 & 1 \end{bmatrix}$ |
| State observer gain | $L_o = [0.4631 \ -1.4336]^T$ |

To determine the optimal control input $u(k)$, which can satisfy the aforementioned input constraint, one should solve the optimization problem in Eq. (12). Considering the other design parameters as $L = \begin{bmatrix} 1 & 0 \\ 0 & 1 \end{bmatrix}$, $R = 1$ and $\epsilon = 0.001$, the general problem will be solved online. In most cases, it is supposed that the changes in the system's parameters are imposed by the operation point of the system. As a general case, consider the switching signal between different piecewise linear dynamic systems independently. Hence, for generating a general condition of representing the system dynamics at any specified time step, two different switching signals are used; including the defined switching signal (DSWS) and random switching signal (RSWS) (Fig. 1).

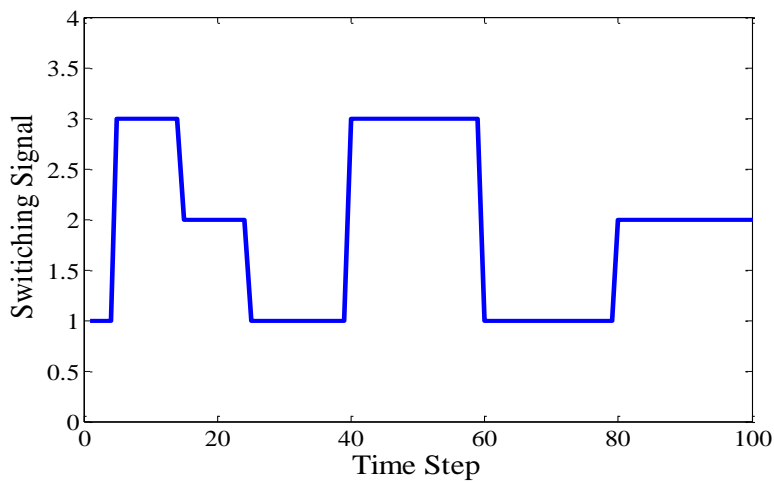

(a)

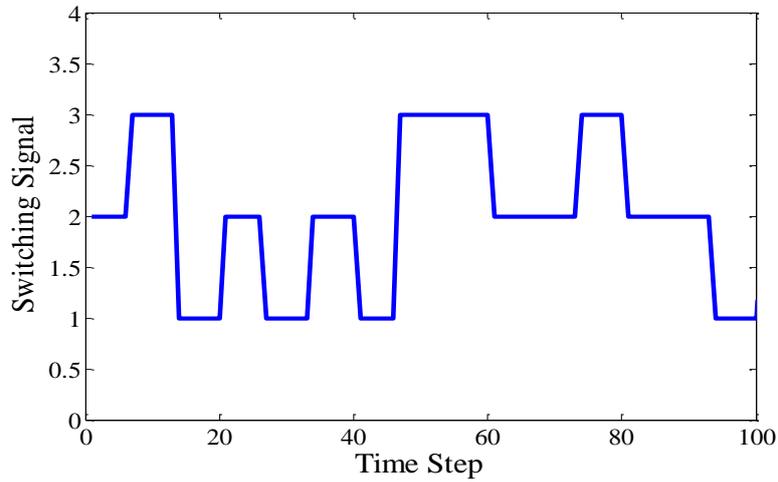

(b)

**Fig. 1** The representation of a dynamic system by the use of switching signals: (a) defined switching signal (DSWS) (b) random switching signal (RSWS)

## 4. RESULTS AND DISCUSSIONS

The aforementioned method was implemented on an LPV system which demonstrates the switching characteristics in its dynamics. Figure 2 shows a comparison of the system output for the proposed improved quasi-min-max approach with output feedback model predictive control versus the one described in the research [31]. The root mean squares (RMS) of the system output for [31] and the current modified approach are 0.1729 and 0.1749, respectively. Despite the introduction of the new cost function for modifying the switching effects in the control input, which are not satisfactory for all systems; the results reveal that both methods have an effective behavior in output regulation (since the proposed modified approach has a slightly larger RMS value). The root mean square (RMS) of the system output for the proposed improved MPC with output feedback versus the one in [31], is presented in Fig. 3 for a given switching signal.

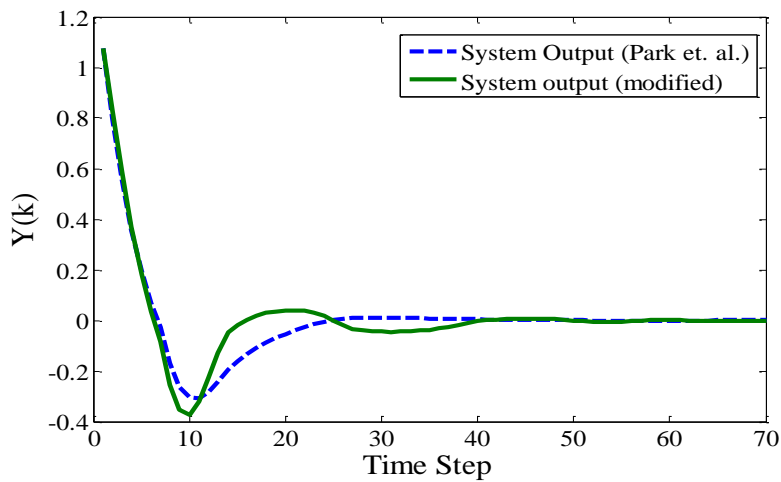

(a)

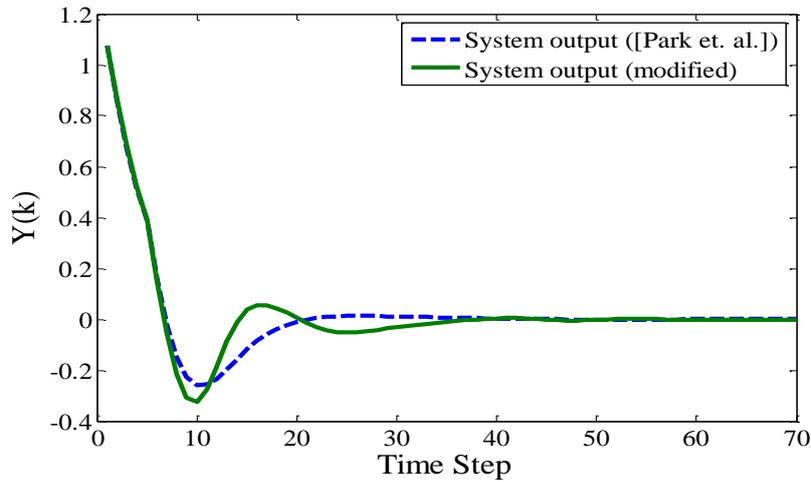

(b)

**Fig. 2** System output; a comparison between the improved approach vs. 31 for $Q = 1 \times I$ and $R = 1$, (a) DSWS (b) RSWS

The presented results in this figure show that by increasing the parameter $Q$ in a constant $R$ ($R = 1$), the reported RMS will not change significantly. Then, for designing the controller, the values for the weighting matrices $R$ and $Q$ are selected as $Q = 1 \times I$ and $R = 1$.

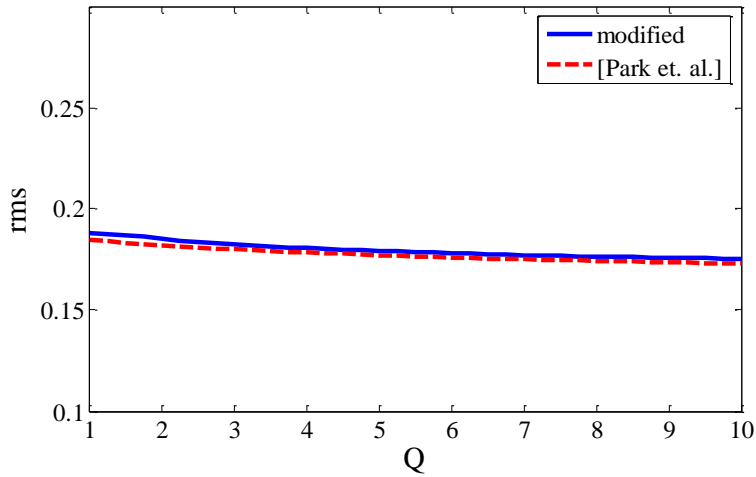

**Fig. 3** The root mean square (RMS) of the system output; a comparison between the improved approach vs. 31, for the variation of $Q$ and $R = 1$ for DSWS

As it is observed in the results, despite the fundamental changes in the definition of the cost function to reduce or even disappear the unwanted switching in the control input; both approaches have similar behavior in regulating the outputs of the control system. A comparison between the estimation errors of the system output for both approaches is presented in Fig. 4. Although both methods have good convergence for tracking the estimated output values, the proposed approach shows faster convergence with little deviation even at the beginning times. Fig. 4 shows that the proposed observer works well for both approaches.

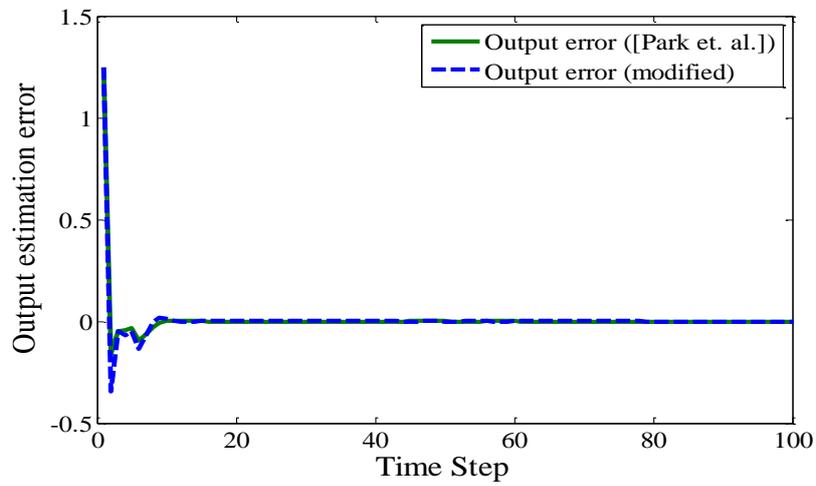

(a)

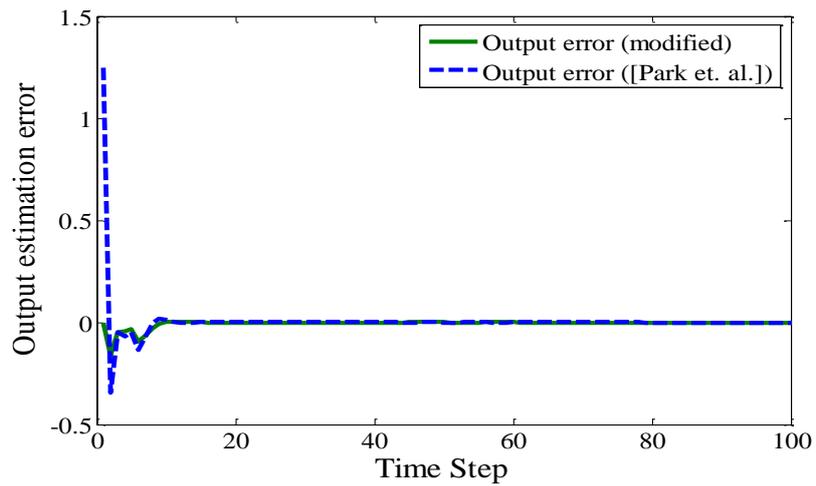

(b)

**Fig. 4** Estimation error of the system output; a comparison between the improved approach vs. 31 for $Q = 1 \times I$ and $R = 1$, (a) DSWS (b) RSWS

Figure 5 shows the applied control input $u(k)$, which is obtained by solving the optimization problem, described by Eq. (44) at each time step. In Fig. 5, the arrows show the points at which the switching occurs in the dynamics of the system. It can be concluded that both methods, the current improved approach and the one in (Park, et al., 2011), have the ability to satisfy the input constraints $|u(k)| < 1; k \geq 0$. It is shown that the proposed approach has better behavior in terms of facing the input fluctuations caused by switches in the dynamics of the system. Also, the initial control effort remains zero at $t = 0$.

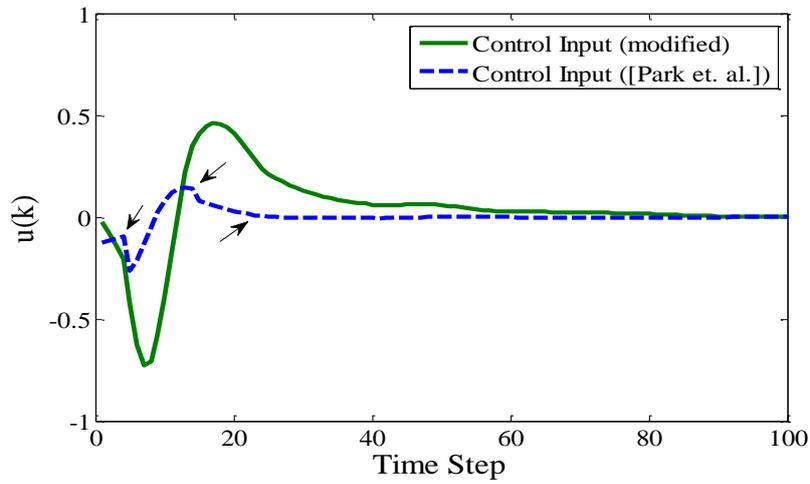

(a)

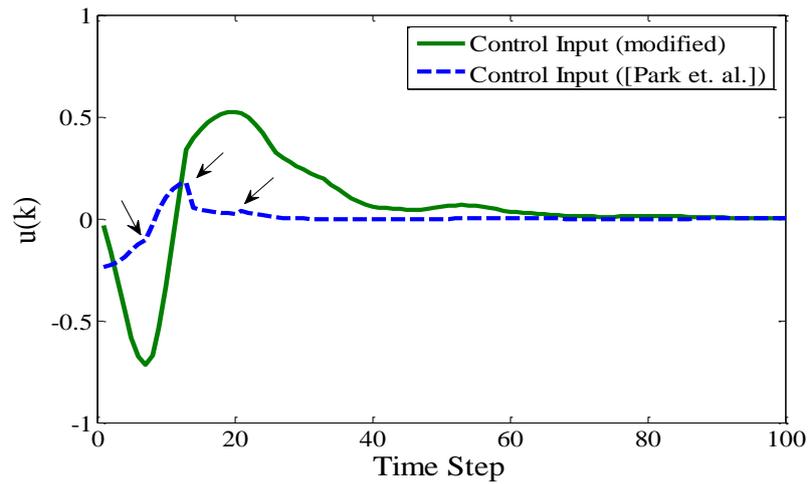

(b)

**Fig. 5** The control input of the system; a comparison between the improved approach vs. 31, for $Q = 1 \times I$ and $R = 1$, (a) DSWS (b) RSWS. Arrows show the points when the switches occur in the dynamics of the system

It also enables to reduce the system uncertainties and improves tracking behavior. However, the switching phenomenon is more dominant when the random switches occur between different linear dynamics of the system (Fig. 5-b).

Figure 6 shows the results of control inputs over time while the values of the weighing parameters have been changed. In this condition, the proposed modified approach of the current research shows a better performance than [31]. As it is observed in Fig. 6, the control inputs can demonstrate the effectiveness of the proposed method. While the proposed method guarantees that the initial value for the control input remains zero and minimizes the sudden changes in the signal due to the switches in the system dynamics. In this regard, the proposed method shows a relatively better performance compared to [31].

In fact, this phenomenon is due to the reduction of uncertainties' effects that results in the system's robustness by the MPC rule; by modifying the control law and also making the weight matrixes independent of the undefined parameters. Hence, it is possible to reduce the unwanted switches and discontinuities in the control efforts and also guarantee the initial value of the control signal to be zero at $t = 0$.

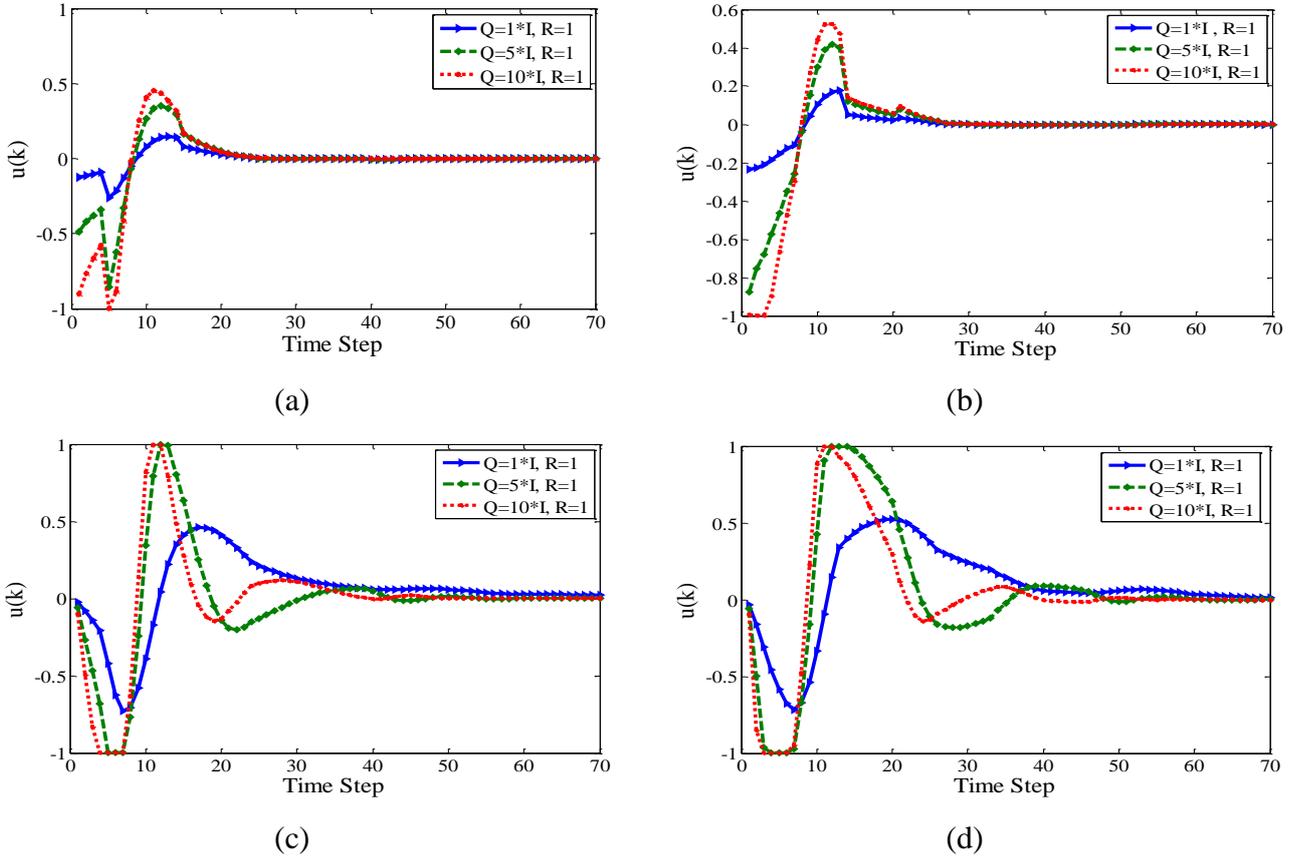

**Fig. 6** Comparison of the system control input; (a) DSWS of 31, (b) RSWS of 31, (c) DSWS of the modified algorithm, (d) RSWS of modified algorithm

## 5. CONCLUSION

In this paper, a modified robust MPC algorithm with output feedback for LPV or hybrid systems is proposed using a quasi-min-max method. First, an offline robust state observer was designed. Then a new cost function was defined for the control design, which has the ability to reduce or even eliminate unwanted disturbances in the control signal. While the robustness of the system is guaranteed, the effect of uncertainties is reduced in the predictive dynamic model. This leads to the convergence of the output signals to the desired values. Moreover, the initial value of the control effort starts with the zero value, which is worthy and important in practice.

The proposed algorithm was implemented on a hybrid system using two different scenarios; one with a defined and the other with a random switching signal in the dynamics. In both cases, while the stability of the whole system and the performance of the designed observer were guaranteed, the control inputs had an initial zero value. In addition, the undesired disturbances in the previous algorithm caused by switching the dynamics were eliminated. In conclusion, the proposed modified

MPC scheme for linear parametric variable or hybrid systems can guarantee the robust stability of the output feedback systems and the convergence of optimization subject to input constraints.

**Conflict of interest:** The authors declare no conflict of interests.